\documentclass[prd,preprint,superscriptaddress]{revtex4}
\usepackage{revsymb}
\usepackage{amsmath}
\newcommand{\be}{\begin{equation}}
\newcommand{\ee}{\end{equation}}
\newcommand{\ben}{\begin{eqnarray}}
\newcommand{\een}{\end{eqnarray}}

\begin{document}
\title{Holographic dark energy and cosmic coincidence}
\author{Diego Pav\'{o}n\footnote{E-mail address: diego.pavon@uab.es}}
\address{Departamento de F\'{\i}sica, Universidad Aut\'{o}noma de
Barcelona, 08193 Bellaterra (Barcelona), Spain}
\author{Winfried Zimdahl\footnote{E-mail address: zimdahl@thp.uni-koeln.de}}
\address{Institut f\"ur Theoretische Physik, Universit\"at zu  K\"oln\\
D-50937 K\"oln, Germany}

\begin{abstract}
In this Letter we demonstrate that any interaction of pressureless
dark matter with holographic dark energy, whose infrared cutoff is
set by the Hubble scale, implies a constant ratio of the energy
densities of both components thus solving the coincidence problem.
The equation of state parameter is obtained as a function of the
interaction strength. For a variable degree of saturation of the
holographic bound the energy density ratio becomes time dependent
which is compatible with a transition from decelerated to
accelerated expansion.
\end{abstract}
\maketitle

Nowadays there is a wide consensus among cosmologists that the
Universe has entered a phase of accelerated expansion
\cite{Riess}. The debate is now centered on  when the acceleration
did actually begin, whether it is to last forever or it is just a
transient episode and, above all, which is the agent behind it.
Whatever the agent, usually called dark energy, it must possess a
negative pressure high enough to violate the strong energy
condition. A number of dark energy candidates have been put
forward, ranging from an incredibly tiny cosmological constant to
a variety of exotic fields (scalar, tachyon, k-essence, etc) with
suitably chosen potentials \cite{reviews}. Most of the candidates,
however, suffer from the coincidence problem, namely: {\em Why are
the matter and dark energy densities of precisely the same order
today?} \cite{coincidence}.

Recently, a new dark energy candidate, based not in any specific
field but on the holographic principle, was proposed
\cite{horava,enqvist,Hsu,mli1,q-gh,mli2}. The latter, first
formulated by 't Hooft \cite{hooft} and Susskind \cite{susskind},
has attracted much attention as a possible short cut to quantum
gravity and found interesting applications in cosmology -see e.g.
\cite{qappl}- and black hole growth \cite{ernesto}. According to
this principle, the number of degrees of freedom of physical
systems scales with their bounding area rather than with their
volume. In this context Cohen {\em et al.} reasoned that the dark
energy should obey the aforesaid principle and be constrained by
the infrared (IR) cutoff \cite{cohen}. In line with this
suggestion, Li has argued that the dark energy density should
satisfy the bound $\rho_{X} \leq 3\, M_{p}^{2}\, c^{2}/L^{2}$,
where $c^2$ is a constant and $M_{p}^{2} = (8\pi\, G)^{-1}$
\cite{mli1}. He discusses three choices for the length scale $L$
which is supposed to provide an IR cutoff. The first choice is to
identify $L$ with the Hubble radius, $H^{-1}$. Applying arguments
from Hsu \cite{Hsu}, Li demonstrates that this leads to a wrong
equation of state, namely that for dust. The second option is the
particle horizon radius. However, this does not work either since
it is impossible to obtain an accelerated expansion on this basis.
Only the third choice, the identification of $L$ with the radius
of the future event horizon gives the desired result, namely a
sufficiently negative equation of state to obtain an accelerated
universe.

Here, we point out that Li's conclusions rely on the assumption of
an independent evolution of the energy densities of dark energy
and matter which, in particular, implies a scaling $\rho_{M}
\propto a^{-3}$ of the matter energy density $\rho_{M}$ with the
scale factor $a(t)$. Any interaction between both components will
change, however, this dependence. The target of this Letter is to
demonstrate that as soon as an interaction is taken into account,
the first choice, the identification of $L$ with $H^{-1}$, can
simultaneously drive accelerated expansion and solve the
coincidence problem. We believe that models of late acceleration
that do not solve the coincidence problem cannot be deemed
satisfactory (see, however, \cite{McInnes}).

Let us reconsider the argument Li used to discard the
identification of the IR cutoff with Hubble's radius. Setting $L =%
H^{-1}$ in the above bound and working with the equality (i.e.,
assuming that the holographic bound is saturated) it becomes
$\rho_{X} = 3\, c^{2}M^{2}_{P}H^{2}$. Combining the last
expression with Friedmann's equation for a spatially flat
universe, $ 3M^{2}_{P}H^{2} = \rho_{X} + \rho_{M}$, results in $
\rho_{M} = 3\left(1 - c^{2}\right)M^{2}_{P}H^{2}$. Now, the
argument runs as follows: The energy density $\rho_{M}$ varies as
$H^2$, which coincides with the dependence of $\rho_{X}$ on $H$.
The energy density of cold matter is known to scale as $\rho_{M}
\propto a^{-3}$. This corresponds to an equation of state $p_{M}
\ll \rho_{M}$, i.e., dust. Consequently, this should be the
equation of state for the dark energy as well. Thus, the dark
energy behaves as pressureless matter. Obviously, pressureless
matter cannot generate accelerated expansion, which seems to rule
out the choice $L=H^{-1}$. This is exactly Li's conclusion. What
underlies this reasoning is the assumption that $\rho_{M}$ and
$\rho_{X}$ evolve independently. However if one realizes that the
ratio of the energy densities
\\
\begin{equation}\label{r}
r \equiv \frac{\rho_{M}}{\rho_{X}} = \frac{1 -%
c^{2}}{c^{2}}  \, ,
\end{equation}
\\
should approach a constant, finite value $r = r_{0}$ for the
coincidence problem to be solved, a different interpretation is
possible, which no longer relies on an independent evolution of
the components. Given the unknown nature of both dark matter and
dark energy there is nothing in principle against their mutual
interaction (however, in order not to conflict with ``fifth force"
experiments \cite{fithf} we do not consider baryonic matter) to
the point that assuming no interaction at all is not less
arbitrary than assuming a coupling. In fact, this possibility is
receiving growing attention in the literature
\cite{plb,CJPZ,interaction} and  appears to be compatible not only
with SNIa and CMB data \cite{german} but even favored over
non-interacting cosmologies \cite{szy}. On the other hand, the
coupling should not be seen as an entirely phenomenological
approach as different Lagrangians have been proposed in support of
the coupling -see \cite{tsu} and references therein.

As a consequence of their mutual interaction  neither component
conserves separately,
\\
\be \dot{\rho}_{M} + 3H \rho_{M} = Q \, , \qquad
\quad\dot{\rho}_{X} + 3H (1+w)\rho_{X} = -Q \, , \label{nconsv}
\ee
\\
though the total energy density, $\rho = \rho_{M}+\rho_{X}$, does.
Here $Q$ denotes the interaction term, and $w$  the equation of
state parameter of the dark energy. Without loss of generality we
shall describe the interaction as a decay process with $Q = \Gamma
\rho_{X}$ where $\Gamma$ is an arbitrary (generally variable)
decay rate. Then we may write
\ \\
\begin{equation}
\dot{\rho}_{M} + 3H \rho_{M} = \Gamma \rho_{X} \label{dotrhom}
\end{equation}
and
\begin{equation}
\dot{\rho}_{X} + 3H (1+w)\rho_{X} = - \Gamma \rho_{X} \, .
\label{dotrhox}
\end{equation}
\ \\
Consequently, the evolution of $r$ is governed by
\\
\begin{equation}
\dot{r} = 3 H r \left[w + \frac{1 + r}{r}\frac{\Gamma}{3H}\right]
\ .\label{dotr}
\end{equation}
\\
In the non-interacting case ($\Gamma = 0$) and for a constant
equation of state parameter $w$ this ratio scales as $r \propto a
^{3 w}$.  If we now assume $ \rho_{X} = 3\, c^{2}M^{2}_{P}H^{2}$,
this definition implies
\\
\begin{equation}
\dot{\rho}_{X}  = - 9\, c^{2}M^{2}_{P} H^3 \, \left[1 + \frac{w}{1
+ r}\right] \ ,\label{dotrhox1}
\end{equation}
\ \\
where we have employed  Einstein's equation $ \dot{H}  = -
\frac{3}{2}\, H^{2}\left[1 + \frac{w}{1 + r}\right]$. Inserting
(\ref{dotrhox1}) in the left hand side of the balance equation
(\ref{dotrhox}) yields a relation between the equation of state
parameter $w$ and the interaction rate $\Gamma$, namely,
\\
\begin{equation}
w = - \left(1 + \frac{1}{r}\right) \frac{\Gamma}{3 H} \, .
\label{Gamma}
\end{equation}
\ \\
The interaction parameter $\frac{\Gamma}{3 H}$ together with the
ratio $r$ determine the equation of state. In the absence of
interaction, i.e., for $\Gamma = 0$, we have $w = 0$, i.e., Li's
result is recovered as a special case. For the choice $ \rho_{X} =
3\, c^{2}M^{2}_{P}H^{2}$ an interaction is the only way to have an
equation of state different from that for dust. Any decay of the
dark energy component ($\Gamma> 0$) into pressureless matter is
necessarily accompanied by an equation of state $w<0$.

\noindent The existence of an interaction has another interesting
consequence. Using the expression (\ref{Gamma}) for $\Gamma$ in
(\ref{dotr}) provides us with $ \dot{r}  = 0$, i.e., $r = r_{0} =$
constant.   Therefore, if the dark energy is given by $ \rho_{X} =
3\, c^{2}M^{2}_{P}\, H^{2}$ and if an interaction with a
pressureless component is admitted, the ratio $r =
\rho_{M}/\rho_{X} $ is necessarily constant, irrespective of the
specific structure of the interaction. \noindent Under this
condition we have [cf.(\ref{r})]
\\
\begin{equation}
 c^{2} = \frac{1}{1+r_{0}}\, .
 \label{r-c}
\end{equation}
\\
At variance with \cite{mli1,mli2}, the fact that $c^{2}$ is lower
than unity does not prompt any conflict with thermodynamics. For
the case of a constant interaction parameter $\frac{\Gamma}{3H}
\equiv \nu =$ constant, it follows that
\\
\begin{equation}\label{rhoprop}
\rho ,\, \rho_{M} ,\, \rho_{X} \propto a^{-3 m} \, \, \qquad \left(m%
= 1 + \frac{w}{1 + r_{0}} = 1 - \frac{\nu}{r}\right)\ ,
\end{equation}
\\
while the scale factor obeys $a \propto t^{n}$ with $n = 2/(3m)$.
Consequently, the condition for accelerated expansion is
$w/(1+r_{0}) < -1/3$, i.e.,  $ \nu > r_{0}/3$.

Accordingly, the expression for the holographic dark energy with
the identification $L = H^{-1}$ fits well into the interacting
dark energy concept. The Hubble radius is not only the most
obvious but also the simplest choice.  It is not only compatible
with a constant ratio between the energy densities but requires
it. In a sense, the holographic dark energy with $L=H^{-1}$
together with the observational fact of an accelerated expansion
almost calls for an interacting model. Note that the interaction
is essential to simultaneously solve the coincidence problem and
have late acceleration. There is no non-interacting limit, since
in the absence of interaction, i.e., $Q = \Gamma = 0$,  there is
no acceleration.

Obviously, a change of $r_{0}$ demands a corresponding change of
$c^{2}$. Within the framework discussed so far, a dynamical
evolution of the energy density ratio is impossible. As a way out
it has been suggested again to replace the Hubble scale by the
future event horizon \cite{Wangetal}. Here we shall follow a
different strategy to admit a dynamical energy density ratio.
Motivated by the relation (\ref{r-c}) in the stationary case $r =
r_{0} =$const, we retain the expression $ \rho_{X} = 3\,
c^{2}M^{2}_{P}H^{2}$ for the dark energy but allow the so far
constant parameter $c^{2}$ to vary, i.e., $c^{2} = c^{2}(t)$.
Since the precise value of $c^{2}$ is unknown, some time
dependence of this parameter cannot be excluded. Then this
definition of $ \rho_{X} $ implies
\\
\begin{equation}
\dot{\rho}_{X}  = - 9\, c^{2}M^{2}_{P} H^3 \, \left[1 + \frac{w}{1
+ r}\right] + \frac{\left(c^{2}\right)^{\displaystyle
\cdot}}{c^{2}}\rho_{X}\ ,\label{dotrhox2}
\end{equation}
\ \\
which generalizes Eq. (\ref{dotrhox1}). Using now the expression
(\ref{dotrhox2}) for $\dot{\rho}_{X}$ on the left hand side of the
balance equation (\ref{dotrhox}), leads to
\\
\begin{equation}
\frac{\left(c^{2}\right)^{\displaystyle \cdot}}{c^{2}} = - 3 H
\frac{r}{1 + r}\left[w + \frac{1 + r}{r}\frac{\Gamma}{3H}\right]
\, .
\label{cdot}
\end{equation}
\ \\
A vanishing left hand side, i.e., $c^{2} = $constant, consistently
reproduces (\ref{Gamma}). Comparing the right hand sides of
equations (\ref{cdot}) and (\ref{dotr}) yields
$\left(c^{2}\right)^{\displaystyle \cdot}/c^{2} = -%
\dot{r}/(1+r)$, whose solution is
\\
\begin{equation}
c^{2} \left(1 + r\right) = 1 \,.
\label{c21+r}
\end{equation}
\\
The constant has been chosen to have the correct behavior
(\ref{r-c}) for the limit $r = r_{0} =$ constant. We conclude that
if the dark energy is given by $\rho_{X} = 3\,c^{2}M^{2}_{P}H^{2}$
and $c^{2}$ is allowed to be time dependent, this time dependence
must necessarily preserve the quantity $c^{2} \left(1 + r\right)$.
The time dependence of $c^{2}$ thus fixes the dynamics of $r$ (and
vice versa). Since $r$ is expected to decrease in the course of
cosmic expansion, $\dot{r} < 0$, this is accompanied by an
increase in $c^{2}$, i.e.,
$\left(c^{2}\right)^{\displaystyle\cdot} > 0$.

Solving (\ref{cdot}) for the equation of state parameter $w$ we
find
\\
\begin{equation}
w = - \left(1 +  \frac{1}{r}\right) \left[\frac{\Gamma}{3 H} +
\frac{\left(c^{2}\right)^{\displaystyle \cdot}}{3 H c^{2}}\right]\
.\label{WX}
\end{equation}
\\
For $\left(c^{2}\right)^{\displaystyle \cdot} =0$ one recovers
expression (\ref{Gamma}). It is obvious, that both a decreasing
$r$ and an increasing $c^2$ in (\ref{WX}) tend to make $w$ more
negative compared with $w = - \left(1 +  \frac{1}{r}\right)
\frac{\Gamma}{3 H}$  from (\ref{Gamma}). A variation of the $c^2$
parameter can be responsible for a change in the  equation of
state parameter $w$. Such a change to (more) negative values is
required for the transition from decelerated to accelerated
expansion. For a specific dynamic model assumptions about the
interaction have to be introduced. This may be done, e.g., along
the lines of \cite{plb,CJPZ}. However, as is well known, the
holographic energy must fulfill the dominant energy condition
\cite{bak} whereby it is not compatible with a phantom equation of
state ($w < -1$). This automatically sets a constraint on $\Gamma$
and $c^{2}$.

It is noteworthy that in allowing $c^{2}$ to vary, contrary to
what one may think, the infrared cutoff does not necessarily
change. This may be  be seen as follows. The holographic bound can
be written as $\rho_{X} \leq 3 c^2 M_{p}^{2}/L^{2} $ with $L =
H^{-1}$. Now, Li and Huang \cite{mli1,q-gh,mli2} -as well  as
ourselves- assume that the holographic bound is saturated (i.e.,
the equality sign is assumed in the above expression). Since the
saturation of the bound is not at all compelling, and the
``constant" $c^2 (t)$ increases with expansion  (as $r$ decreases)
up to reaching the constant value $(1+ r_{0})^{-1} $, the
expression $ \rho_{X} = 3 c^2 (t) M_{p}^{2} H^{2}$, in reality,
does not imply a modification of the infrared cutoff, which is
still $L = H^{-1}$. What happens is that, as $c^{2}(t)$ grows, the
bound gets progressively saturated up to full saturation when,
asymptotically, $c^{2}$ becomes a constant. In other words, the
infrared cutoff always remains $L = H^{-1}$, what changes is the
degree of saturation of the holographic bound.

In this letter we have shown that {\it any} interaction of a dark
energy component with density $\rho_{X} = 3\, c^{2}M^{2}_{P}H^{2}$
(and $c^2 =$ constant) with a pressureless dark matter component
necessarily implies a constant ratio of the energy densities of
both components. The equation of state parameter $w$ is determined
by the interaction strength. A time evolution of the energy
density ratio is uniquely related to a time variation of the
$c^{2}$ parameter. Under this condition a  decreasing ratio
$\rho_{M}/\rho_{X}$ sends $w$ to lower values.

\end{document}